\documentclass[pre,floatfix,twocolumn,showkeys,a4paper,superscriptaddress,footinbib,showpacs]{revtex4-1}
\usepackage{amsmath}
\usepackage{amsfonts}
\usepackage{mathrsfs}
\usepackage{epsfig}
\usepackage{graphicx}
\usepackage{amsmath}
\usepackage{color}

\begin{document}

\title{Using explosive percolation in analysis of  real-world networks}

\author{Raj Kumar Pan}
\author{Mikko Kivel\"a}
\author{Jari Saram\"aki}
\author{Kimmo Kaski}
\affiliation{BECS, Aalto University School of Science, P.O. Box 12200, FI-00076}
\author{J\'anos Kert\'esz}
\affiliation{Institute of Physics and HAS-BME Cond.~Mat.~Group, BME, Budapest, Budafoki \'ut 8., H-1111}
\affiliation{BECS, Aalto University School of Science, P.O. Box 12200, FI-00076}

\date{\today}

\begin{abstract}
We apply a variant of the explosive percolation procedure to large
real-world networks, and show with finite-size scaling that the university
class, ordinary or explosive, of the resulting percolation transition
depends on the structural properties of the network as well as the number
of unoccupied links considered for comparison in our procedure.  We observe
that in our social networks, the percolation clusters close to the critical
point are related to the community structure. This relationship is further
highlighted by applying the procedure to model networks with pre-defined
communities.

\end{abstract}


\maketitle

\section{Introduction}
The percolation process realized by the {\em Achlioptas
procedure}~\cite{Achlioptas09} is different from classical percolation.
This ``explosive percolation'' begins with a graph of isolated nodes and at
each step, 
two potential edges are chosen at random. Then, the edge that minimizes the
product or sum 
of the sizes of the two components that would be merged is added to the
graph. This procedure eventually leads to an explosive percolation
transition that appears discontinuous (first order). However, it
has recently been argued that in reality the transition is continuous and belongs to a
new universality class with very small exponent of the order
parameter~\cite{daCosta10}. The above or similar procedures have been
applied to various model networks ranging from regular lattices~\cite{Ziff09}
to scale-free networks~\cite{Radicchi09,*Cho09}. Several papers have
painted an intuitive picture of the mechanisms behind this behavior
such as local cluster aggregation~\cite{Cho10,*DSouza10}, formation of many
large components before percolation transition~\cite{Friedman09}, or
inhibition of growth of the largest cluster~\cite{Araujo10}. Other criteria
for the growth process have also been suggested, such as choosing edges
proportionally to a weight determined by their cluster
sizes~\cite{Manna11}. 

While explosive percolation has triggered a considerable amount of
theoretical and simulation work, its application to real-world networks or
processes has been limited~\cite{Rozenfeld10}. The topological
characteristics of real-world networks, such as high clustering, degree
correlations, community structure, and weight-topology correlations, are far from those of 
regular or random model graphs~\cite{Newman06a}. Such
features play a role in the characteristics of classical percolation that
has earlier been successfully applied to investigate real-world network
structure. Here we ask if they also play a crucial role in explosive
percolation, and if monitoring the percolation process itself yields
important information about the network structure. As a pre-requisite, we establish that
proper link addition rules yield explosive percolation transitions when
applied to real-world networks. However, this depends both on the network
structure and the details of the evolution rules.

\section{Data and methods}

For our empirical networks, we have chosen a mobile phone call network
(MPC)~\cite{Onnela07} and a large ArXiv co-authorship network
(CA)~\cite{Newman01a}. Both networks are social, so that nodes
represent people and ties their interactions, and are large enough for
percolation studies. They also share features common to social networks,
such as community structure and assortativity~\cite{Newman06a}. For the
MPC, it has been shown that tie strengths relate to network
topology: strong ties are associated with dense network neighborhoods
(communities)~\cite{Granovetter73, *Onnela07b}. Such weight-topology
correlations are reflected in classical percolation behavior. For the CA,
to the best of our knowledge, weight-topology correlations have not been
studied in detail before.

The MPC data consists of $325\times10^6$ voice calls over a period of 120
days. We construct an aggregated undirected weighted network of edges
with bidirectional calls between users, weights representing the total
number of calls. The largest connected component (LCC) is then extracted,
with $4.6 \times 10^6$ nodes and $9.1 \times 10^6$ edges.  The
collaboration data is from the arXiv~\cite{Arxiv} and contains
all e-prints in  ``physics'' until March 2010. There are
 4.8 $\times$ $10^5$ article headers, from which we
extract the authors.  In the co-authorship (CA) network two authors are
connected if they have co-authored articles, whose number determines the
link weight. We then extract the LCC, with  $1.8 \times 10^5$ nodes and
$9.1 \times 10^6$ edges.  In addition, we
construct a filtered version of the CA, where articles with more than 10
authors ($\sim 2\%$ of articles) are ignored. This is to remove the
very large cliques from papers with $\sim 10^3$ authors in fields
such as hep-ex or astro-ph, where the principles behind collaboration
network formation appear different.  The LCC of the resulting small
collaboration co-authorship (SCA) network has $1.5 \times 10^5$ nodes and
$9.1 \times 10^5$ edges. Note that, although the number of nodes is not much
smaller than for the CA, the number of edges is an order of magnitude less.

For the percolation process, we use the min-cluster (MC-$m$) sum rule with
different values of $m$, defined as follows. Initially, all the edges of the
empirical network are considered unoccupied. Then, at each time step, $m$
unoccupied edges are drawn at random. Out of these, the edge that would
minimize the size of the component formed if the edge were occupied is
chosen. Intra-component edges are always favored against inter-component
edges as they do not increase the size of any cluster. When comparing two
inter-component edges, we select the one for which the sum of cluster sizes
that it connects is minimized. Ties are resolved randomly. We also study
the limiting case ($m=\infty$), where all unoccupied edges are considered
at each step. This leads to a semi-deterministic process where all
intra-cluster links get occupied before the cluster grows in size. The only
source of randomness is the existence of clusters of same size during the
process~\footnote{A similar rule has been introduced independently in J.S.
Andrade, {\em et.al.,} arXiv:1010.5097, which appear after submission of
this manuscript.}.

\section{Results}

\subsection{Percolation analysis}

\begin{figure*}[ht!]
  \begin{center}
    \includegraphics[width=0.8\linewidth]{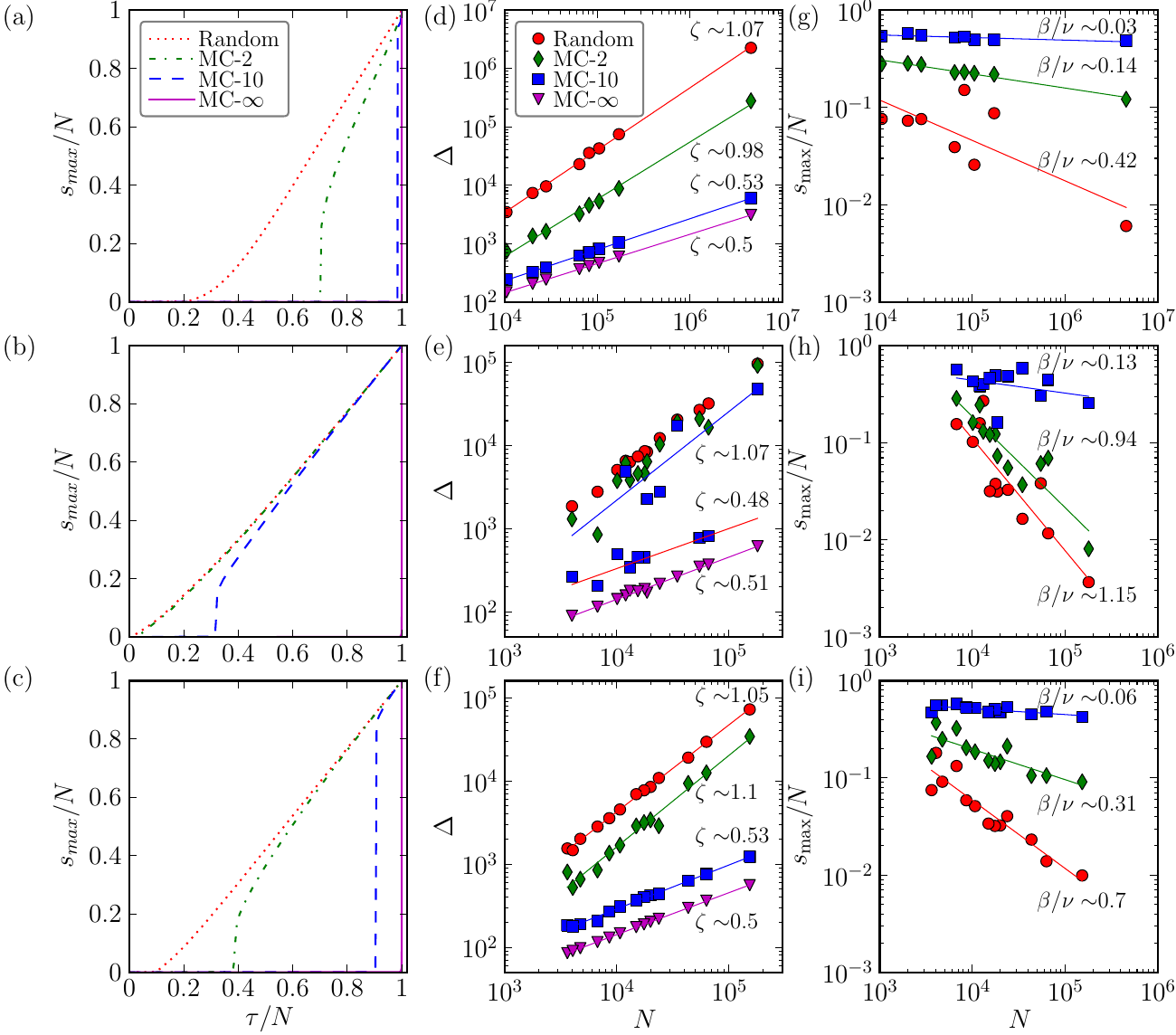}
  \end{center}
  \caption{Variation of the relative size of giant component,
  $s_{\textrm{max}}/N$, with scaled number of inter-cluster edges $\tau/N$
  for the (a) MPC (b) CA and (c) SCA network. The corresponding variations
  in the gap, $\Delta \equiv \tau(N/2)-\tau(\sqrt{N})$, as a function of
  system sizes are shown in (d), (e) and (f), for the Random, MC-2, MC-10
  and MC-$\infty$ rules. Solid lines indicate fitted scaling exponents
  $\zeta$.  The variation of the order parameter, $s_{max}/N$ as a function
  of the system size $N$ is shown for (g) MPC (h) CA and (i) SCA network.
  For each system the order parameter is calculated at the critical point.
  The solid line indicates the best fit obtained and the exponent
  $\beta/\nu$. All curves are averaged over $10^3$ runs. 
  }
  \label{fig:gcc_w_tau}
\end{figure*}

Let us first monitor the behavior of the order parameter, \emph{i.e.} the
relative size of the largest cluster, $s_{\mathrm{max}}/N$, as the fraction
of occupied edges $f_{\mathrm{links}}$ is increased.  As intra-cluster
edges do not affect cluster growth,  we consider the number
of inter-cluster edges $\tau$ instead of $f_{\mathrm{links}}$~\footnote{
In literature, $f_{\mathrm{links}}$ has also been used as the control
parameter, with slightly different MC rules; for those, intra-cluster links
do play a role~\cite{Moreira10}.}.
We apply
three variants of the MC rule: MC-2, MC-10, and MC-$\infty$, as well as
random link percolation for comparison. Fig.~\ref{fig:gcc_w_tau}~(a,b,c)
shows the variation of the fraction $s_{\mathrm{max}} (\tau)/N$ against the
scaled number of inter-component edges, $\tau/N$. For all three networks,
the transition of the order parameter is smooth for the random case, while for
the extreme case, MC-$\infty$, the transition appears abrupt. However, for
MC-2 and MC-10, the situation is more complicated, and we study them in
detail.

\begin{figure}
  \begin{center}
    \includegraphics[width=1.0\linewidth]{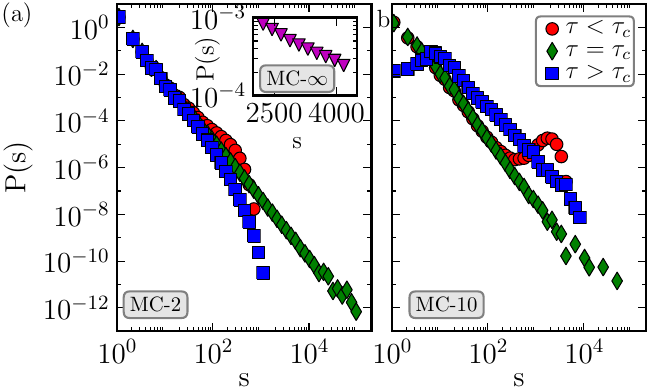}
  \end{center}
  \caption{Cluster size distributions around the critical $\tau_c$ for the MPC, for MC-2 (a), MC-10 (b), and MC-$\infty$ (inset).}
  \label{fig:sizeDist}
\end{figure}

To determine the nature of the transition, Achlioptas \emph{et
al.}~\cite{Achlioptas09} studied the dependence of the width of the
transition window on system size. This width can be
quantified as $\Delta \equiv \tau(N/2)-\tau(\sqrt{N})$, where $\tau(N/2)$
and $\tau(\sqrt{N})$ are the lowest values of $\tau$ for which
$s_{\mathrm{max}} > N/2$ and $s_{\mathrm{max}} > \sqrt{N}$, respectively.
In general the width scales as a power law with the system size,  $\Delta
\propto N^\zeta$. For classical percolation, $\zeta=1$. It was argued that for
explosive percolation  $\zeta<1$ and the rescaled width of the transition
region, $\Delta/N \propto N^{\zeta-1}$, vanishes in the limit of large $N$.
While recent results~\cite{daCosta10} argue that the transition region
is in reality finite, the very small exponent of the order parameter
guarantees that in practice it is vanishingly small even for large systems.

For applying finite-size scaling to empirical networks,
samples of different sizes are needed. In general, unbiased sampling of a
network is difficult. Here, we take advantage of the known properties
of our networks.  Call networks are geographically
embedded~\cite{Krings09, *AhnLinkCommunitiesNature2010}, and we
extract subnetworks of users in chosen cities, based on postal codes of
their subscriptions. For the co-authorship networks, we extract sub-networks
of authors with articles in the same subject
class. We see that for all networks $\Delta \tau
\propto N^\zeta$, with $\zeta\sim1$ for random and $\zeta\sim0.5$ for the
MC-$\infty$ case (Fig.~\ref{fig:gcc_w_tau}~(d,e,f)). Thus the exponent
$\zeta$ clearly differentiates the explosive transition from
random-link percolation. Further, for all three networks, $\zeta\sim1$ for the
MC-2, resembling an ordinary percolation transition. However, for MC-10,
the scaling exponent behaves differently for the three networks. For the MPC
and SCA networks, $\zeta\sim0.5$, indicating explosive percolation. For the
CA, at first it appears that the data points do not follow scaling.
However, a closer inspection shows that they cluster around two straight
lines with $\zeta \sim 1$ and $\zeta \sim 0.5$. Indeed, for
subnetworks with large collaborations (e.g., hep-ex, hep-ph) 
$\zeta\sim1$, whereas for other subject classes (e.g., cond-mat, math-ph),
$\zeta\sim0.5$.

In addition, we have performed a finite-size scaling analysis
of the order parameter $s_{\mathrm{max}}/N$~\cite{Radicchi10}.
The scaling relation for $s_{\mathrm{max}}/N$ is given by
\begin{equation}
  \frac{s_{\mathrm{max}}}{N} = N^{-\beta/\nu}F[(\tau-\tau_c)N^{1/\nu}],
\end{equation}
where $F$ is some universal function, $\tau$ is the control parameter,
$\tau_c$ is the critical point of transition, $\beta$ is the critical
exponent of the order parameter and $\nu $ that of the correlation length.
We choose the critical value $\tau_c$ of the control parameter as the value of  $\tau$ 
where the susceptibility, i.e., average cluster size has its maximum. Note that
$\tau_c$ could also be chosen as the point where the cluster size distribution 
becomes a power law~\cite{daCosta10}; however, since our range of 
network sizes includes fairly small networks, this would be too inaccurate as
in some cases there are not enough clusters for determining the shape of 
the distribution.

For the mobile phone call (MPC) network~[Fig~\ref{fig:gcc_w_tau}~(g)], we find
that the scaling at $\tau_c$ of the order parameter $s_{\mathrm{max}}/N$
yields a very small exponent $\beta/\nu \sim 0.03$ for the MC-10 case, while
for MC-2 and random percolation, the exponents are larger,
$\beta/\nu \sim 0.14$ and $\beta/\nu \sim 0.42$, respectively. The
exponents for the small collaboration co-authorship (SCA) network behave
similarly~[Fig~\ref{fig:gcc_w_tau}~(h)], with a low value $\beta/\nu \sim 0.06$ for the
MC-10 case, and relatively high values $\beta/\nu \sim 0.31$ and $\beta/\nu \sim
0.70$ for MC-2 and random percolation, respectively. In contrast, for
the co-authorship (CA) network, the exponents have high values for all
cases~[Fig~\ref{fig:gcc_w_tau}~(i)], $\beta/\nu \sim 0.13$, $\beta/\nu \sim
0.94$ and $\beta/\nu \sim 1.15$, for MC-10, MC-2 and random percolation,
respectively.
 
In order to compare our results to the existing literature, we follow the
relations for the critical exponents given in Ref~\cite{daCosta10}:
$\beta/\nu = \beta/(4\beta+1)$. The value for the exponent $\beta
\sim 0.0555$ given in Ref~\cite{daCosta10} yields $\beta/\nu \sim 0.0455$. 
This value is consistent with
our observation that the transition for MC-10 is explosive in the MPC and
SCA networks, while it is ordinary in the CA network. Note that such small
but finite values of the exponent are consistent with a 2nd order
transition; however, because we are dealing with single, finite-size
networks, we cannot make definite conclusions.  Further, in all the three
systems MC-2 behaves similar to the ordinary random percolation.

Thus our percolation analysis on CA and SCA networks reveals a difference
between collaboration structures in different fields.  One possible
explanation is the broad degree distribution for the CA network, whose tail
can be approximated with a power law with exponent $\sim 1.7$ in contrast
to SCA, which decays as $\sim 4.3$.  Hence, in this respect, the SCA
network structure resembles the social network of the MPC. Further, it is
clear that the nature of the transition depends both on the number of edges
$m$ considered in the percolation process and structural features of the
network.

\begin{figure}
  \begin{center}
    \includegraphics[width=1.0\linewidth]{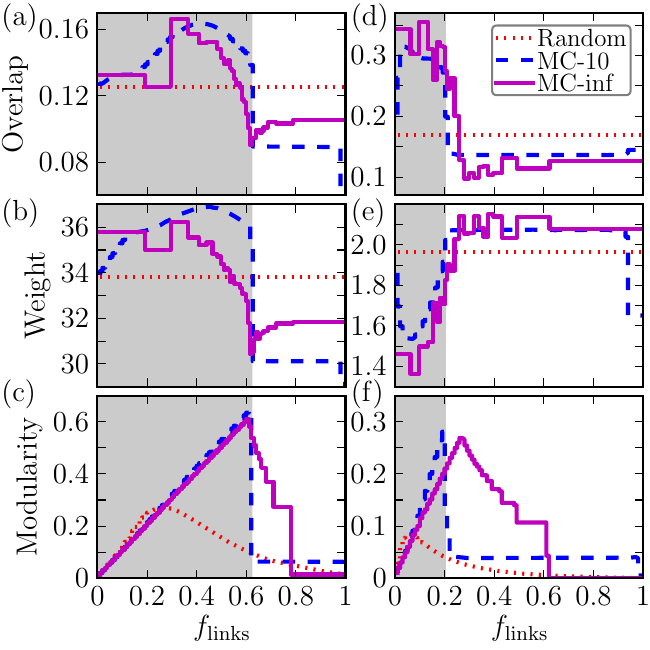}
  \end{center}
  \caption{Variation of the overlap, edge weight, and modularity as
  a function of the fraction of links added,  for MPC (a,b,c) and SCA
  (d,e,f). The shaded area denotes the non-percolating regime for MC-10.}
  \label{fig:edgeProp_phone}
\end{figure}

For the rest of this paper we focus only on the complete MPC and SCA networks, and
first study their cluster size distributions around the critical point,
$\tau_c$.  
For the following, we have chosen $\tau_c$ as the point at which $P(s)$ is a power law
for the full region of $s$~\cite{daCosta10}. The complete networks are large enough
to choose $\tau_c$ this way, giving us in this case a more 
precise value than the susceptibility peaks. Then, we sweep the value of $\tau$ around this point
and monitor the distribution of cluster sizes.
Fig.~\ref{fig:sizeDist}~(a,b) shows the cluster size distributions $P(s)$
around $\tau_c$ for the MCP, for  MC-2, MC-10, and MC-$\infty$. For MC-2,
$P(s)$ behaves as usual for ordinary percolation, becoming a power law at
$\tau_c$ and then turning exponential.  For MC-10, the situation is
different: for $\tau<\tau_c$, there is a bump in the tail of the
distribution, in line with theoretical predictions for explosive
percolation~\cite{daCosta10}.  Immediately above $t_c$, the smallest
remaining clusters get depleted from the distribution as they are the first
to join the giant cluster. For the semi-deterministic MC-$\infty$ (inset),
the cluster size distribution resembles exponential for $\tau<\tau_c$. The
cluster size distributions for SCA are qualitatively similar. 

\subsection{Percolation clusters,  weight-topology correlations, and communities}

Next, we investigate the evolution of the percolation clusters and their
relationship to communities and the weight-topology correlations. We study
the overlap of the neighborhoods of endpoint nodes $i$ and $j$ of a link,
defined as 
\begin{equation}
  O_{ij} = n_{ij}/(k_{i}-1+k_{j}-1-n_{ij}),
\end{equation}
where $n_{ij}$ is the number of neighbors common to both nodes, and $k_i$
and $k_j$ are their degrees~\cite{Onnela07}. This measure quantifies the extent by which
two connected nodes share their neighborhoods: if $i$ and $j$ have no common neighbors, then
 $O_{ij}=0$, and if $i$ and $j$ share all of their neighbors, $O_{ij}=1$. 
 Thus if there are dense communities in the network, links inside the communities
 have high values of overlap, whereas links acting as ``bridges'' connecting separate communities
 have low overlap values.

Fig.~\ref{fig:edgeProp_phone}~(a) displays the
results for the MPC network. As expected, for random link addition, the
overlap and the time when edges are added in the percolation process
are uncorrelated. For MC-10 and MC-$\infty$, edges with high overlap and weight
are added first [Fig.~\ref{fig:edgeProp_phone}~(a,b)]. This indicates that
dense regions of the network, \emph{i.e.} communities, get percolated
first.  Both quantities
show an abrupt drop at the transition point. This fits well with the
Granovetterian weight-topology correlations observed
earlier~\cite{Onnela07}.  However, the behavior of the SCA network is
different. Although high-overlap edges are added first
[Fig.~\ref{fig:edgeProp_phone}~(d)], their weights are low
[Fig.~\ref{fig:edgeProp_phone}~(e)]. This points towards fundamentally
different weight-topology correlations, where strong links act as bridges
between communities of weaker links. A likely explanation is that
communities organize around senior scientists (hubs), with whom junior
researchers are linked. The latter has a small number of joint
publications with the local hubs, as they are only temporarily connected.
The hubs, in turn, are linked via long-lasting
collaborations and many co-authored papers. 

The relationship to
community structure is confirmed with the behavior of the
modularity~\cite{Newman04a} of percolation clusters, defined as  
\begin{equation}
  \mathcal M=\sum_{c}[(L_c/L)-(d_c/2L)^2],
\end{equation}
where the sum runs over clusters, $L$ is the number of links in the
network, $L_c$ is the number of links within cluster $c$, and $d_{c}$ is
the sum of the degrees of nodes in $c$. High values of $\mathcal M$
correspond to a good community
partition -- hence, a high value of modularity
calculated for percolation clusters indicates that they match well with 
communities. As for the other quantities, we calculate $\mathcal M$ as a function
of the fraction of links added $f_{\mathrm{links}}$. As seen in~[Fig.~\ref{fig:edgeProp_phone}~(e,f)], the peak of $\mathcal M$ 
and the following sharp transition match the transition points well for
MC-10. For the semi-deterministic MC-$\infty$, the peak also matches the
percolation point although the transition is less sharp.

\begin{figure}
  \begin{center}    
    \includegraphics[width=0.49\linewidth]{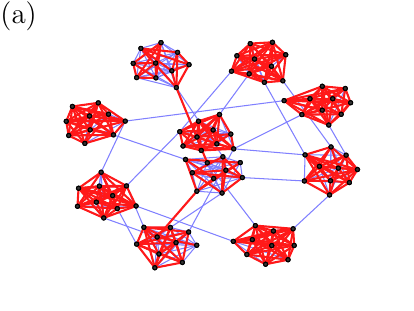}
    \includegraphics[width=0.49\linewidth]{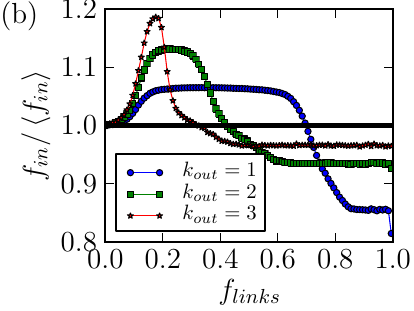}
    \includegraphics[width=0.49\linewidth]{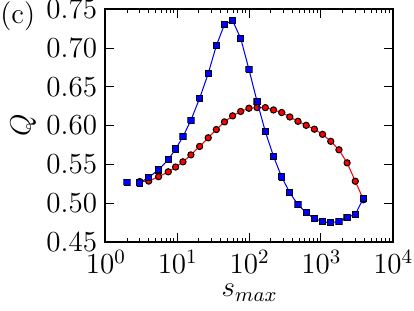}
    \includegraphics[width=0.49\linewidth]{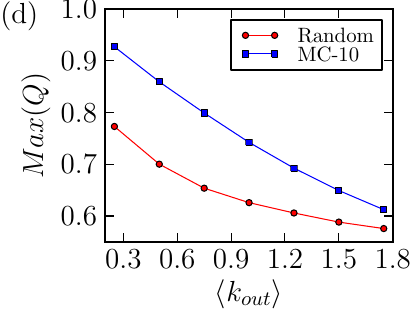}
  \end{center}
  \caption{
  (a) Occupied (red) and unoccupied (blue) edges before the critical point
  in the model network with the MC-10 rule. Here, $N=100$, $M=10$,
  $k_{\mathrm{in}}=9.6$ and $k_{\mathrm{out}}=0.4$. (b) The
  fraction of intra-community links $f_{\mathrm{in}}$ during the
  percolation process normalized by the average fraction $\left<
  f_{\mathrm{in}}\right>$ for random link addition. (c) Matching quality
  $Q$ against largest cluster size, $s_{\mathrm{max}}$ for the model
  network. (d) Maximum of the quality, $Q_{\mathrm{max}}$, as a function of
  $k_{\mathrm{out}}$. All curves are shown for the model with $N=4096$,
  $M=128$ and $k_{\mathrm{out}}+k_{\mathrm{in}}=16$.  For (b) and (c)
  $k_{\mathrm{out}}=1$.}
  \label{fig:gnmodel_overlap}
\end{figure}

\subsection{Analysis of network model with communities}
It appears that the explosive percolation process follows community
structure when applied to a network where such structure exists.
Communities in real-world networks are, however, hard to define
unambiguously, and therefore  we turn to a simple model with built-in
community structure~\cite{Newman04a,Pan09a}. In this model, $N$ nodes are arranged
into $M$ communities of equal size, and edges are placed at random such
that on average each node has $k_{\textrm{in}}$ intra-community links and
$k_{\textrm{out}}$ inter-community links. When applying the MC-$m$ sum rule
to this network, we find that mostly intra-community edges are occupied
before the transition point [Fig.~\ref{fig:gnmodel_overlap}~(a)].  We
quantify this by measuring the fraction of intra-community links that have been added during the
process, normalized by the respective fraction for random link addition. It is evident
from Fig.~\ref{fig:gnmodel_overlap}~(b) that the MC rules prefer intra-community links 
early on in the process, and inter-community links only get added towards the end.

To quantify the match between
percolation clusters and the model communities, we consider the confusion
matrix with elements 
\begin{equation}
n_{kk^{'}} = \lvert C_{k} \cap C^{'}_{k^{'}} \rvert, ~\forall k,k^{'},
\end{equation}
where $C_{k}$ is the $k$-th cluster and $C^{'}_{k^{'}}$
is the $k^{'}$-th community.  Hence, the element $n_{kk^{'}}$ represents
the number of nodes in the intersection of cluster $C_{k}$ and community
$C^{'}_{k^{'}}$.

 There is a perfect match if clusters are subsets of communities
and vice versa, \emph{i.e.} clusters equal communities. 
The extent to which clusters are subsets of communities can be measured by the projection number of $C$
on $C^{'}$, defined as 
\begin{equation}
p_C(C^{'}) = \sum_k \max_{k^{'}} n_{kk^{'}},
\end{equation}
i.e., the sum of the maximum of each row in the confusion matrix.
$p_C(C^{'})$ increases with cluster size, reaching its maximum when there is a single
cluster that overlaps with all communities. For the reverse case, communities as subsets of clusters, one can define a similar projection number
$p_{C^{'}}(C)$, i.e., the sum of the maximum of each column in the
matrix. This number is maximized when the clusters are as small as possible, \emph{i.e.} single nodes,
and decreases with
increasing cluster size~\cite{vanDongen00,*Meila07}.  The {\em quality} of
matching can now be quantified with the normalized average of both
projection numbers, 
\begin{equation}
  Q = [p_C(C^{'})+p_{C^{'}}(C)]/2N,
\end{equation}
reaching its maximum when the match between clusters and communities is
optimal.  

Fig.~\ref{fig:gnmodel_overlap}~(c) shows the behavior of $Q$ for the model network as a function
of the size of the largest observed cluster $s_{\mathrm{max}}$ spanned by the added links. Here we use
the largest cluster size $s_{\mathrm{max}}$ instead of $f_{\mathrm{links}}$ because this provides us 
with a more detailed view on what happens around the transition point; the cluster
sizes change only a little beyond this region.
It is seen that $Q$ initially increases and
then decreases as a function of  $s_{\mathrm{max}}$, 
reaching its maximum before the formation of the giant
component and merging of clusters.
The percolation clusters coincide well with the model communities below and
around $\tau_c$ compared to random link addition.  We next study the
behavior of the maximum of quality $Q_{\mathrm{max}}$ as we make the community
structure more smeared-out by increasing
$k_{\mathrm{out}}$ while keeping the average total degree fixed
[Fig.~\ref{fig:gnmodel_overlap}~(d)]. Although $Q_{\mathrm{max}}$ decreases
as $k_{out}$ increases for both the MC-$10$ and random addition, its higher
value for the MC-$10$ process indicates better match with the built-in
communities. 

We also obtain qualitatively similar results by using normalized the mutual
information (NMI) instead of the matching quality $Q$ (not shown). The
mutual information~\cite{Danon05} can be
defined using the confusion matrix as
\begin{equation}
I(C,C^{'})=\sum_{k,k^{'}}\frac{n_{kk^{'}}}{N}\log\frac{n_{kk^{'}}N}{n_kn_{k^{'}}}, 
\end{equation}
where $n_k=\sum_{k^{'}}n_{kk^{'}}$ and $n_{k^{'}}=\sum_kn_{kk^{'}}$ are the
size of the $k$-th community and $k^{'}$-th cluster, respectively. The
normalized mutual information is then defined as
\begin{equation}
NMI(C,C^{'})=\frac{2I(C,C^{'})}{H(C)+H(C^{'})}, 
\end{equation}
where $H(C)=-\sum_k n_k/N \log(n_k/N)$ is the entropy of the community $C$,
and $H(C^{'})$ is the entropy of the cluster $C^{'}$. In our case, where we
have large number of small communities, the NMI does not however work as well as
the matching quality. This is because the NMI values are high already at the beginning
of the percolation process when all the nodes are isolated forming their
own clusters. In this case, $NMI(C,C^{'})=2(\frac{\log N}{\log M}+1)^{-1}$,
which approaches $1$ if the model network size is increased keeping the
community sizes, $N/M$ fixed. In contrast, the initial value of the quality is
$Q=\frac{1}{2}+\frac{M}{2N}$, which is independent of the number of
communities.

\section{Summary and conclusions}

To summarize, we have shown that the Achlioptas procedure can give rise to
an explosive percolation transition when the rules are applied to empirical
real-world social networks. We have used a variant of the Minimum Cluster
(MC) rule, where the number of links compared during the link addition
process is a parameter, and shown that both the network structure and  the
number of links compared have an influence on the universality class
(ordinary or explosive) of the observed percolation transition.  In order
to show this, we have carried out finite-size scaling using subnetworks,
chosen on the basis of known external properties of the empirical networks.
This is an important but non-trivial task when percolation analysis is
applied to empirical networks where only a single ``realization'' is
available.  The resulting values for critical exponents are in line with
the view that the explosive percolation transition is in fact second order;
however, one cannot make definite conclusions since we are dealing with
singe, finite-size networks.

In addition, we have illustrated a connection between links selected by the MC rule 
during the percolation process and community structure -- at the critical point,
the cluster structure arising from the application of the MC rule reflects the community
structure of the network. This is confirmed by the analysis of single-link properties (the overlap, link weight), 
 and modularity for the empirical networks, and by 
detailed studies of the match between clusters and built-in community structure 
of model networks.

\acknowledgements
Financial support from EU's 7th Framework Program's FET-Open to
ICTeCollective project no. 238597 and by the Academy of Finland, the
Finnish Center of Excellence program 2006-2011, project no. 129670, as well
as by TEKES (FiDiPro) are gratefully acknowledged. We thank A.-L.
Barab\'asi for the MPC data used in this research.

\bibliography{explosive_net.bib}

\begin{thebibliography}{10}%
\makeatletter
\providecommand \@ifxundefined [1]{%
 \ifx #1\undefined \expandafter \@firstoftwo
 \else \expandafter \@secondoftwo
\fi
}%
\providecommand \@ifnum [1]{%
 \ifnum #1\expandafter \@firstoftwo
 \else \expandafter \@secondoftwo
\fi
}%
\providecommand \enquote [1]{``#1''}%
\providecommand \bibnamefont  [1]{#1}%
\providecommand \bibfnamefont [1]{#1}%
\providecommand \citenamefont [1]{#1}%
\providecommand\href[0]{\@sanitize\@href}%
\providecommand\@href[1]{\endgroup\@@startlink{#1}\endgroup\@@href}%
\providecommand\@@href[1]{#1\@@endlink}%
\providecommand \@sanitize [0]{\begingroup\catcode`\&12\catcode`\#12\relax}%
\@ifxundefined \pdfoutput {\@firstoftwo}{%
 \@ifnum{\z@=\pdfoutput}{\@firstoftwo}{\@secondoftwo}%
}{%
 \providecommand\@@startlink[1]{\leavevmode\special{html:<a href="#1">}}%
 \providecommand\@@endlink[0]{\special{html:</a>}}%
}{%
 \providecommand\@@startlink[1]{%
  \leavevmode
  \pdfstartlink
   attr{/Border[0 0 1 ]/H/I/C[0 1 1]}%
   user{/Subtype/Link/A<</Type/Action/S/URI/URI(#1)>>}%
  \relax
 }%
 \providecommand\@@endlink[0]{\pdfendlink}%
}%
\providecommand \url  [0]{\begingroup\@sanitize \@url }%
\providecommand \@url [1]{\endgroup\@href {#1}{\urlprefix}}%
\providecommand \urlprefix [0]{URL }%
\providecommand \Eprint[0]{\href }%
\@ifxundefined \urlstyle {%
  \providecommand \doi [1]{doi:\discretionary{}{}{}#1}%
}{%
  \providecommand \doi [0]{doi:\discretionary{}{}{}\begingroup
  \urlstyle{rm}\Url }%
}%
\providecommand \doibase [0]{http://dx.doi.org/}%
\providecommand \Doi[1]{\href{\doibase#1}}%
\providecommand \bibAnnote [3]{%
  \BibitemShut{#1}%
  \begin{quotation}\noindent
    \textsc{Key:}\ #2\\\textsc{Annotation:}\ #3%
  \end{quotation}%
}%
\providecommand \bibAnnoteFile [2]{%
  \IfFileExists{#2}{\bibAnnote {#1} {#2} {\input{#2}}}{}%
}%
\providecommand \typeout [0]{\immediate \write \m@ne }%
\providecommand \selectlanguage [0]{\@gobble}%
\providecommand \bibinfo [0]{\@secondoftwo}%
\providecommand \bibfield [0]{\@secondoftwo}%
\providecommand \translation [1]{[#1]}%
\providecommand \BibitemOpen[0]{}%
\providecommand \bibitemStop [0]{}%
\providecommand \bibitemNoStop [0]{.\EOS\space}%
\providecommand \EOS [0]{\spacefactor3000\relax}%
\providecommand \BibitemShut [1]{\csname bibitem#1\endcsname}%
\bibitem{Achlioptas09}%
  \BibitemOpen
  \bibfield{author}{%
  \bibinfo {author} {\bibfnamefont{D.}~\bibnamefont{Achlioptas}}, \bibinfo
  {author} {\bibfnamefont{R.~M.}\ \bibnamefont{D'Souza}},\ and\ \bibinfo
  {author} {\bibfnamefont{J.}~\bibnamefont{Spencer}},\ }%
  \bibfield{journal}{%
  \Doi{10.1126/science.1167782}{\bibinfo {journal} {Science}}\ }%
  \textbf{\bibinfo {volume} {323}},\ \bibinfo {pages} {1453} (\bibinfo {year}
  {2009})%
  \bibAnnoteFile{NoStop}{Achlioptas09}%
\bibitem{daCosta10}%
  \BibitemOpen
  \bibfield{author}{%
  \bibinfo {author} {\bibfnamefont{R.~A.}\ \bibnamefont{da~Costa}}, \bibinfo
  {author} {\bibfnamefont{S.~N.}\ \bibnamefont{Dorogovtsev}}, \bibinfo {author}
  {\bibfnamefont{A.~V.}\ \bibnamefont{Goltsev}},\ and\ \bibinfo {author}
  {\bibfnamefont{J.~F.~F.}\ \bibnamefont{Mendes}},\ }%
  \bibfield{journal}{%
  \Doi{10.1103/PhysRevLett.105.255701}{\bibinfo {journal} {Phys. Rev. Lett.}}\
  }%
  \textbf{\bibinfo {volume} {105}},\ \bibinfo {pages} {255701} (\bibinfo {year}
  {2010})%
  \bibAnnoteFile{NoStop}{daCosta10}%
\bibitem{Ziff09}%
  \BibitemOpen
  \bibfield{author}{%
  \bibinfo {author} {\bibfnamefont{R.~M.}\ \bibnamefont{Ziff}},\ }%
  \bibfield{journal}{%
  \Doi{10.1103/PhysRevLett.103.045701}{\bibinfo {journal} {Phys. Rev. Lett.}}\
  }%
  \textbf{\bibinfo {volume} {103}},\ \bibinfo {pages} {045701} (\bibinfo {year}
  {2009})%
  \bibAnnoteFile{NoStop}{Ziff09}%
\bibitem{Radicchi09}%
  \BibitemOpen
  \bibfield{author}{%
  \bibinfo {author} {\bibfnamefont{F.}~\bibnamefont{Radicchi}}\ and\ \bibinfo
  {author} {\bibfnamefont{S.}~\bibnamefont{Fortunato}},\ }%
  \bibfield{journal}{%
  \Doi{10.1103/PhysRevLett.103.168701}{\bibinfo {journal} {Phys. Rev. Lett.}}\
  }%
  \textbf{\bibinfo {volume} {103}},\ \bibinfo {pages} {168701} (\bibinfo {year}
  {2009})%
  \bibAnnoteFile{NoStop}{Radicchi09}%
\bibitem{Cho09}%
  \BibitemOpen
  \bibfield{author}{%
  \bibinfo {author} {\bibfnamefont{Y.~S.}\ \bibnamefont{Cho}}, \bibinfo
  {author} {\bibfnamefont{J.~S.}\ \bibnamefont{Kim}}, \bibinfo {author}
  {\bibfnamefont{J.}~\bibnamefont{Park}}, \bibinfo {author}
  {\bibfnamefont{B.}~\bibnamefont{Kahng}},\ and\ \bibinfo {author}
  {\bibfnamefont{D.}~\bibnamefont{Kim}},\ }%
  \bibfield{journal}{%
  \Doi{10.1103/PhysRevLett.103.135702}{\bibinfo {journal} {Phys. Rev. Lett.}}\
  }%
  \textbf{\bibinfo {volume} {103}},\ \bibinfo {pages} {135702} (\bibinfo {year}
  {2009})%
  \bibAnnoteFile{NoStop}{Cho09}%
\bibitem{Cho10}%
  \BibitemOpen
  \bibfield{author}{%
  \bibinfo {author} {\bibfnamefont{Y.~S.}\ \bibnamefont{Cho}}, \bibinfo
  {author} {\bibfnamefont{B.}~\bibnamefont{Kahng}},\ and\ \bibinfo {author}
  {\bibfnamefont{D.}~\bibnamefont{Kim}},\ }%
  \bibfield{journal}{%
  \Doi{10.1103/PhysRevE.81.030103}{\bibinfo {journal} {Phys. Rev. E}}\ }%
  \textbf{\bibinfo {volume} {81}},\ \bibinfo {pages} {030103} (\bibinfo {year}
  {2010})%
  \bibAnnoteFile{NoStop}{Cho10}%
\bibitem{DSouza10}%
  \BibitemOpen
  \bibfield{author}{%
  \bibinfo {author} {\bibfnamefont{R.~M.}\ \bibnamefont{D'Souza}}\ and\
  \bibinfo {author} {\bibfnamefont{M.}~\bibnamefont{Mitzenmacher}},\ }%
  \bibfield{journal}{%
  \Doi{10.1103/PhysRevLett.104.195702}{\bibinfo {journal} {Phys. Rev. Lett.}}\
  }%
  \textbf{\bibinfo {volume} {104}},\ \bibinfo {pages} {195702} (\bibinfo {year}
  {2010})%
  \bibAnnoteFile{NoStop}{DSouza10}%
\bibitem{Friedman09}%
  \BibitemOpen
  \bibfield{author}{%
  \bibinfo {author} {\bibfnamefont{E.~J.}\ \bibnamefont{Friedman}}\ and\
  \bibinfo {author} {\bibfnamefont{A.~S.}\ \bibnamefont{Landsberg}},\ }%
  \bibfield{journal}{%
  \Doi{10.1103/PhysRevLett.103.255701}{\bibinfo {journal} {Phys. Rev. Lett.}}\
  }%
  \textbf{\bibinfo {volume} {103}},\ \bibinfo {pages} {255701} (\bibinfo {year}
  {2009})%
  \bibAnnoteFile{NoStop}{Friedman09}%
\bibitem{Araujo10}%
  \BibitemOpen
  \bibfield{author}{%
  \bibinfo {author} {\bibfnamefont{N.~A.~M.}\ \bibnamefont{Ara\'ujo}}\ and\
  \bibinfo {author} {\bibfnamefont{H.~J.}\ \bibnamefont{Herrmann}},\ }%
  \bibfield{journal}{%
  \Doi{10.1103/PhysRevLett.105.035701}{\bibinfo {journal} {Phys. Rev. Lett.}}\
  }%
  \textbf{\bibinfo {volume} {105}},\ \bibinfo {pages} {035701} (\bibinfo {year}
  {2010})%
  \bibAnnoteFile{NoStop}{Araujo10}%
\bibitem{Manna11}%
  \BibitemOpen
  \bibfield{author}{%
  \bibinfo {author} {\bibfnamefont{S.}~\bibnamefont{Manna}}\ and\ \bibinfo
  {author} {\bibfnamefont{A.}~\bibnamefont{Chatterjee}},\ }%
  \bibfield{journal}{%
  \Doi{DOI: 10.1016/j.physa.2010.10.009}{\bibinfo {journal} {Physica A}}\ }%
  \textbf{\bibinfo {volume} {390}},\ \bibinfo {pages} {177 } (\bibinfo {year}
  {2011})%
  \bibAnnoteFile{NoStop}{Manna11}%
\bibitem{Rozenfeld10}%
  \BibitemOpen
  \bibfield{author}{%
  \bibinfo {author} {\bibfnamefont{H.~D.}\ \bibnamefont{Rozenfeld}}, \bibinfo
  {author} {\bibfnamefont{L.~K.}\ \bibnamefont{Gallos}},\ and\ \bibinfo
  {author} {\bibfnamefont{H.~A.}\ \bibnamefont{Makse}},\ }%
  \bibfield{journal}{%
  \Doi{10.1140/epjb/e2010-00156-8}{\bibinfo {journal} {Eur. Phys. J. B}}\ }%
  \textbf{\bibinfo {volume} {75}},\ \bibinfo {pages} {305} (\bibinfo {year}
  {2010})%
  \bibAnnoteFile{NoStop}{Rozenfeld10}%
\bibitem{Newman06a}%
  \BibitemOpen
  \bibfield{author}{%
  \bibinfo {author} {\bibfnamefont{M.}~\bibnamefont{Newman}}, \bibinfo {author}
  {\bibfnamefont{A.~L.}\ \bibnamefont{Barabasi}},\ and\ \bibinfo {author}
  {\bibfnamefont{D.~J.}\ \bibnamefont{Watts}},\ }%
  \emph{\bibinfo {title} {The Structure and Dynamics of Networks}}\ (\bibinfo
  {publisher} {Princeton University Press},\ \bibinfo {address} {Princeton},\
  \bibinfo {year} {2006})%
  \bibAnnoteFile{NoStop}{Newman06a}%
\bibitem{Onnela07}%
  \BibitemOpen
  \bibfield{author}{%
  \bibinfo {author} {\bibfnamefont{J.~P.}\ \bibnamefont{Onnela}}, \bibinfo
  {author} {\bibfnamefont{J.}~\bibnamefont{Saram\"aki}}, \bibinfo {author}
  {\bibfnamefont{J.}~\bibnamefont{Hyv\"onen}}, \bibinfo {author}
  {\bibfnamefont{G.}~\bibnamefont{Szab\'o}}, \bibinfo {author}
  {\bibfnamefont{D.}~\bibnamefont{Lazer}}, \bibinfo {author}
  {\bibfnamefont{K.}~\bibnamefont{Kaski}}, \bibinfo {author}
  {\bibfnamefont{J.}~\bibnamefont{Kert\'esz}},\ and\ \bibinfo {author}
  {\bibfnamefont{A.~L.}\ \bibnamefont{Barab\'asi}},\ }%
  \bibfield{journal}{%
  \bibinfo {journal} {Proc. Natl. Acad. Sci. U.S.A.}\ }%
  \textbf{\bibinfo {volume} {104}},\ \bibinfo {pages} {7332} (\bibinfo {year}
  {2007})%
  \bibAnnoteFile{NoStop}{Onnela07}%
\bibitem{Newman01a}%
  \BibitemOpen
  \bibfield{author}{%
  \bibinfo {author} {\bibfnamefont{M.~E.~J.}\ \bibnamefont{Newman}},\ }%
  \bibfield{journal}{%
  \Doi{10.1103/PhysRevE.64.016132}{\bibinfo {journal} {Phys. Rev. E}}\ }%
  \textbf{\bibinfo {volume} {64}},\ \bibinfo {pages} {016132} (\bibinfo {year}
  {2001})%
  \bibAnnoteFile{NoStop}{Newman01a}%
\bibitem{Granovetter73}%
  \BibitemOpen
  \bibfield{author}{%
  \bibinfo {author} {\bibfnamefont{M.}~\bibnamefont{Granovetter}},\ }%
  \bibfield{journal}{%
  \bibinfo {journal} {Am. J. Soc.}\ }%
  \textbf{\bibinfo {volume} {78}},\ \bibinfo {pages} {1360} (\bibinfo {year}
  {1973})%
  \bibAnnoteFile{NoStop}{Granovetter73}%
\bibitem{Onnela07b}%
  \BibitemOpen
  \bibfield{author}{%
  \bibinfo {author} {\bibfnamefont{J.-P.}\ \bibnamefont{Onnela}}, \bibinfo
  {author} {\bibfnamefont{J.}~\bibnamefont{Saram\"aki}}, \bibinfo {author}
  {\bibfnamefont{J.}~\bibnamefont{Hyv\"onen}}, \bibinfo {author}
  {\bibfnamefont{G.}~\bibnamefont{Szab\'o}}, \bibinfo {author}
  {\bibfnamefont{M.~A.}\ \bibnamefont{de~Menezes}}, \bibinfo {author}
  {\bibfnamefont{K.}~\bibnamefont{Kaski}}, \bibinfo {author}
  {\bibfnamefont{A.-L.}\ \bibnamefont{Barab\'asi}},\ and\ \bibinfo {author}
  {\bibfnamefont{J.}~\bibnamefont{Kert\'esz}},\ }%
  \bibfield{journal}{%
  \bibinfo {journal} {New J. Phys.}\ }%
  \textbf{\bibinfo {volume} {9}},\ \bibinfo {pages} {179} (\bibinfo {year}
  {2007})%
  \bibAnnoteFile{NoStop}{Onnela07b}%
\bibitem{Arxiv}%
  \BibitemOpen
  \bibinfo {author} {\bibnamefont{http://arxiv.org/}}%
  \bibAnnoteFile{NoStop}{Arxiv}%
\bibitem{Note1}%
  \BibitemOpen
\bibfield{author}{%
    }%
  \bibinfo {note} {A similar rule has been introduced independently in J.S.
  Andrade, {\protect \em et.al.,} arXiv:1010.5097, which appear after
  submission of this manuscript.}%
  \bibAnnoteFile{Stop}{Note1}%
\bibitem{Note2}%
  \BibitemOpen
  \bibinfo {note} {In literature, $f_{\protect \mathrm {links}}$ has also been
  used as the control parameter, with slightly different MC rules; for those,
  intra-cluster links do play a role~\cite {Moreira10}.}%
  \bibAnnoteFile{Stop}{Note2}%
\bibitem{Krings09}%
  \BibitemOpen
  \bibfield{author}{%
  \bibinfo {author} {\bibfnamefont{G.}~\bibnamefont{Krings}}, \bibinfo {author}
  {\bibfnamefont{F.}~\bibnamefont{Calabrese}}, \bibinfo {author}
  {\bibfnamefont{C.}~\bibnamefont{Ratti}},\ and\ \bibinfo {author}
  {\bibfnamefont{V.~D.}\ \bibnamefont{Blondel}},\ }%
  \bibfield{journal}{%
  \bibinfo {journal} {J. Stat. Mech.}\ }%
  \textbf{\bibinfo {volume} {2009}},\ \bibinfo {pages} {L07003} (\bibinfo
  {year} {2009})%
  \bibAnnoteFile{NoStop}{Krings09}%
\bibitem{AhnLinkCommunitiesNature2010}%
  \BibitemOpen
  \bibfield{author}{%
  \bibinfo {author} {\bibfnamefont{Y.-Y.}\ \bibnamefont{Ahn}}, \bibinfo
  {author} {\bibfnamefont{J.~P.}\ \bibnamefont{Bagrow}},\ and\ \bibinfo
  {author} {\bibfnamefont{S.}~\bibnamefont{Lehmann}},\ }%
  \bibfield{journal}{%
  \bibinfo {journal} {Nature}\ }%
  \textbf{\bibinfo {volume} {466}},\ \bibinfo {pages} {761} (\bibinfo {year}
  {2010})%
  \bibAnnoteFile{NoStop}{AhnLinkCommunitiesNature2010}%
\bibitem{Radicchi10}%
  \BibitemOpen
  \bibfield{author}{%
  \bibinfo {author} {\bibfnamefont{F.}~\bibnamefont{Radicchi}}\ and\ \bibinfo
  {author} {\bibfnamefont{S.}~\bibnamefont{Fortunato}},\ }%
  \bibfield{journal}{%
  \Doi{10.1103/PhysRevE.81.036110}{\bibinfo {journal} {Phys. Rev. E}}\ }%
  \textbf{\bibinfo {volume} {81}},\ \bibinfo {pages} {036110} (\bibinfo {year}
  {2010})%
  \bibAnnoteFile{NoStop}{Radicchi10}%
\bibitem{Newman04a}%
  \BibitemOpen
  \bibfield{author}{%
  \bibinfo {author} {\bibfnamefont{M.~E.~J.}\ \bibnamefont{Newman}}\ and\
  \bibinfo {author} {\bibfnamefont{M.}~\bibnamefont{Girvan}},\ }%
  \bibfield{journal}{%
  \Doi{http://link.aps.org/abstract/PRE/v69/e026113}{\bibinfo {journal} {Phys.
  Rev. E}}\ }%
  \textbf{\bibinfo {volume} {69}},\ \bibinfo {eid} {026113} (\bibinfo {year}
  {2004})%
  \bibAnnoteFile{NoStop}{Newman04a}%
\bibitem{Pan09a}%
  \BibitemOpen
  \bibfield{author}{%
  \bibinfo {author} {\bibfnamefont{R.~K.}\ \bibnamefont{Pan}}\ and\ \bibinfo
  {author} {\bibfnamefont{S.}~\bibnamefont{Sinha}},\ }%
  \bibfield{journal}{%
  \bibinfo {journal} {Europhys. Lett.}\ }%
  \textbf{\bibinfo {volume} {85}},\ \bibinfo {pages} {68006} (\bibinfo {year}
  {2009})%
  \bibAnnoteFile{NoStop}{Pan09a}%
\bibitem{vanDongen00}%
  \BibitemOpen
  \bibfield{author}{%
  \bibinfo {author} {\bibfnamefont{S.}~\bibnamefont{Van~Dongen}},\ }%
  \emph{\bibinfo {title} {Performance criteria for graph clustering and Markov
  cluster experiments}},\ \bibinfo {type} {Technical Report}\ \bibinfo {number}
  {INS-R0012}\ (\bibinfo {institution} {Center for Mathematics and Computer
  Science (CWI)},\ \bibinfo {address} {Amsterdam},\ \bibinfo {year} {2000})%
  \bibAnnoteFile{NoStop}{vanDongen00}%
\bibitem{Meila07}%
  \BibitemOpen
  \bibfield{author}{%
  \bibinfo {author} {\bibfnamefont{M.}~\bibnamefont{Meila}},\ }%
  \bibfield{journal}{%
  \Doi{10.1016/j.jmva.2006.11.013}{\bibinfo {journal} {J. Multivariate
  Analysis}}\ }%
  \textbf{\bibinfo {volume} {98}},\ \bibinfo {pages} {873} (\bibinfo {year}
  {2007})%
  \bibAnnoteFile{NoStop}{Meila07}%
\bibitem{Danon05}%
  \BibitemOpen
  \bibfield{author}{%
  \bibinfo {author} {\bibfnamefont{L.}~\bibnamefont{Danon}}, \bibinfo {author}
  {\bibfnamefont{A.}~\bibnamefont{Diaz-Guilera}}, \bibinfo {author}
  {\bibfnamefont{J.}~\bibnamefont{Duch}},\ and\ \bibinfo {author}
  {\bibfnamefont{A.}~\bibnamefont{Arenas}},\ }%
  \bibfield{journal}{%
  \bibinfo {journal} {J. Stat. Mech.},\ \bibinfo {pages} {P09008}}%
   (\bibinfo {month} {September}\ \bibinfo {year} {2005})%
  \bibAnnoteFile{NoStop}{Danon05}%
\bibitem{Moreira10}%
  \BibitemOpen
  \bibfield{author}{%
  \bibinfo {author} {\bibfnamefont{A.~A.}\ \bibnamefont{Moreira}}, \bibinfo
  {author} {\bibfnamefont{E.~A.}\ \bibnamefont{Oliveira}}, \bibinfo {author}
  {\bibfnamefont{S.~D.~S.}\ \bibnamefont{Reis}}, \bibinfo {author}
  {\bibfnamefont{H.~J.}\ \bibnamefont{Herrmann}},\ and\ \bibinfo {author}
  {\bibfnamefont{J.~S.}\ \bibnamefont{Andrade}},\ }%
  \bibfield{journal}{%
  \Doi{10.1103/PhysRevE.81.040101}{\bibinfo {journal} {Phys. Rev. E}}\ }%
  \textbf{\bibinfo {volume} {81}},\ \bibinfo {pages} {040101} (\bibinfo {year}
  {2010})%
  \bibAnnoteFile{NoStop}{Moreira10}%
\end{thebibliography}%
\end{document}